\documentclass[utf8]{frontiersSCNS} 

\usepackage[hyphens]{url}
\usepackage{lineno,microtype,subcaption,caption,graphicx,tabularx,booktabs}
\usepackage[hidelinks]{hyperref}

\usepackage[onehalfspacing]{setspace}

\usepackage{refstyle}
\newref{fig}{refcmd=\hyperref[#1]{Figure~\ref{#1}}}
\newref{tab}{refcmd=\hyperref[#1]{Table~\ref{#1}}}
\newref{sec}{refcmd=\hyperref[#1]{Section~\ref{#1}}}
\newref{eq}{refcmd=\hyperref[#1]{Eq.~\ref{#1}}}
\newref{alg}{refcmd=\hyperref[#1]{Algorithm~\ref{#1}}}
\newref{lst}{refcmd=\hyperref[#1]{Listing~\ref{#1}}}

\usepackage{xcolor}

\usepackage{units}

\usepackage{xcolor}
\definecolor{mymaroon}{cmyk}{0, 0.87, 0.68, 0.32}
\definecolor{myhalfgray}{gray}{0.55}
\definecolor{myframe}{RGB}{207, 207, 207}
\definecolor{mybg}{RGB}{250, 250, 250}
\definecolor{myred}{RGB}{186, 33, 33}
\definecolor{mygreen}{RGB}{0, 128, 0}
\definecolor{mycyan}{RGB}{64, 128, 128}
\definecolor{mypurple}{RGB}{170, 34, 255}
\definecolor{myorange}{RGB}{206, 92, 0}
\definecolor{myblue}{RGB}{32, 74, 135}

\usepackage{inconsolata}

\usepackage{listings}
\lstdefinelanguage{myYAML}{
    morekeywords={parameterset,name,parameter,type,_},
    keywordstyle=\color{mygreen}\ttfamily\textbf,
    morecomment=[l]\#,
    commentstyle=\color{mycyan}\ttfamily\textit,
    morestring=[b]",
    stringstyle=\color{myred}\ttfamily,
}
\lstdefinelanguage{myPython}{
    morekeywords={nest},
    morekeywords=[2]{False},
    keywordstyle=\color{mygreen}\ttfamily\textbf,
    keywordstyle=[2]\color{myblue}\ttfamily\textbf,
	morestring=[b]",
    stringstyle=\color{myred}\ttfamily,
    framexrightmargin=0mm,
}

\usepackage{algorithm2e}
\SetAlgorithmName{Listing}{}{}

\makeatletter
\newcommand\footnoteref[1]{\protected@xdef\@thefnmark{\ref{#1}}\@footnotemark}
\makeatother

\def\keyFont{\fontsize{8}{11}\helveticabold }
\def\firstAuthorLast{Albers {et~al.}} 
\def\Authors{
Jasper Albers\,$^{1,2*}$,
Jari Pronold\,$^{1,2}$,
Anno Christopher Kurth\,$^{1,2}$,
Stine Brekke Vennemo\,$^{3}$,
Kaveh Haghighi Mood\,$^{4}$,
Alexander Patronis\,$^{4}$,
Dennis Terhorst\,$^{1}$,
Jakob Jordan\,$^{5}$,
Susanne Kunkel\,$^{3}$,
Tom Tetzlaff\,$^{1}$,
Markus Diesmann\,$^{1,6,7}$,
and Johanna Senk\,$^{1}$}
\def\Address{
$^{1}$Institute of Neuroscience and Medicine (INM-6) and Institute for Advanced Simulation (IAS-6) and JARA-Institute Brain Structure-Function Relationships (INM-10), Jülich Research Centre, Jülich, Germany \\
$^{2}$RWTH Aachen University, Aachen, Germany \\
$^{3}$Faculty of Science and Technology, Norwegian University of Life Sciences, Ås, Norway\\
$^{4}$Jülich Supercomputing Centre (JSC), Jülich Research Centre, Jülich, Germany \\
$^{5}$Department of Physiology, University of Bern, Bern, Switzerland \\
$^{6}$Department of Physics, Faculty 1, RWTH Aachen University, Aachen, Germany \\
$^{7}$Department of Psychiatry, Psychotherapy and Psychosomatics, School of Medicine, RWTH Aachen University, Aachen, Germany
}
\def\corrAuthor{Jasper Albers}

\def\corrEmail{j.albers@fz-juelich.de}

\begin{document}
\onecolumn
\firstpage{1}

\title[\texttt{beNNch}: Benchmarking Neuronal Network Simulations]{A Modular Workflow for Performance Benchmarking of Neuronal Network Simulations} 

\author[\firstAuthorLast ]{\Authors} 
\address{} 
\correspondance{} 

\extraAuth{}

\maketitle

\begin{abstract}
\section{}
Modern computational neuroscience strives to develop complex network models to explain dynamics and function of brains in health and disease.
This process goes hand in hand with advancements in the theory of neuronal networks and increasing availability of detailed anatomical data on brain connectivity.
Large-scale models that study interactions between multiple brain areas with intricate connectivity and investigate phenomena on long time scales such as system-level learning require progress in simulation speed.
The corresponding development of state-of-the-art simulation engines relies on information provided by benchmark simulations which assess the time-to-solution for scientifically relevant, complementary network models using various combinations of hardware and software revisions.
However, maintaining comparability of benchmark results is difficult due to a lack of standardized specifications for measuring the scaling performance of simulators on high-performance computing (HPC) systems.
Motivated by the challenging complexity of benchmarking, we define a generic workflow that decomposes the endeavor into unique segments consisting of separate modules.
As a reference implementation for the conceptual workflow, we develop \texttt{beNNch}: an open-source software framework for the configuration, execution, and analysis of benchmarks for neuronal network simulations.
The framework records benchmarking data and metadata in a unified way to foster reproducibility.
For illustration, we measure the performance of various versions of the \texttt{NEST} simulator across network models with different levels of complexity on a contemporary HPC system, demonstrating how performance bottlenecks can be identified, ultimately guiding the development toward more efficient simulation technology.
\tiny
 \keyFont{ \section{Keywords:} spiking neuronal networks, benchmarking, large-scale simulation, high-performance computing, workflow, metadata} 
\end{abstract}

\section{Introduction}
\label{sec:introduction}
%
Past decades of computational neuroscience have achieved a separation between mathematical models and generic simulation technology \citep{Einevoll19_735}. This enables researchers to simulate different models with the same simulation engine, while the efficiency of the simulator can be incrementally advanced and maintained as a research infrastructure. Increasing computational efficiency does not only decrease the required resources of simulations, but also allows for constructing larger network models with an extended explanatory scope and facilitates studying long-term effects such as learning. Novel simulation technologies are typically published together with verification\textemdash evidence that the implementation returns correct results\textemdash and validation\textemdash evidence that these results are computed efficiently. Verification implies correctness of results with sufficient accuracy for suitable applications as well as a flawless implementation of components confirmed by unit tests. For spiking neuronal network simulators, such applications are simulations of network models which have proven to be of relevance for the field. In a parallel effort, validation aims at demonstrating the added value of the new technology for the community. To this end, the new technology is compared to previous studies on the basis of relevant performance measures.

Efficiency is measured by the resources used to achieve the result. Time-to-solution, energy-to-solution and memory consumption are of particular interest. For the development of neuromorphic computing systems, efficiency in terms of low power consumption and fast execution is an explicit design goal:
simulations need to be able to cope with limited resources, for example, due to hardware constraints. Real-time performance, meaning that simulated model time equals wall-clock time, is a prerequisite for simulations interacting with the outer world, such as in robotics. Even faster, sub-real-time simulations enable studies of slow neurobiological processes such as brain development and learning, which take hours, days, or more in nature. High-performance computing (HPC) benchmarking studies usually assess the scaling performance of the simulation architecture by incrementally increasing the amount of employed hardware resources (e.g., compute nodes). In weak-scaling experiments, the size of the simulated network model is increased proportionally to the computational resources, which keeps the work load per compute node fixed if the simulation scales perfectly.
Scaling neuronal networks, however, inevitably leads to changes in the network dynamics \citep{Albada15}. Comparisons between benchmarking results obtained at different scales are therefore problematic. 
For network models of natural size describing the correlation structure of neuronal activity, strong-scaling experiments (in which the model size remains unchanged) are more relevant for the purpose of finding the limiting time-to-solution. For a formal definition of strong and weak scaling refer to page 123 of \cite{Hager10_978} and for pitfalls in interpreting the scaling of network simulation code see \cite{Albada14}.
When measuring time-to-solution, studies distinguish between different phases of the simulation, in the simplest case between a setup phase of network construction and the actual simulation phase of state propagation. Such benchmark metrics not only depend on the simulation engine and its options for time measurements \citep[see, e.g.,][]{Jordan18_2,Golosio21_627620}, but also on the network model. The simulated activity of a model may not always be stationary over time, and transients with varying firing rates are reflected in the computational load. For an example of transients due to arbitrary initial conditions, see \cite{Rhodes19_20190160}, and for an example of non-stationary network activity, refer to the meta-stable state of the multi-area model described by \cite{Schmidt18_1409}. Studies assessing energy-to-solution need to specify whether only the power consumption of the compute nodes is considered or interconnects and required support hardware are also accounted for \citep{VanAlbada18_291}.

\begin{figure}[t!]
\begin{center}
\includegraphics{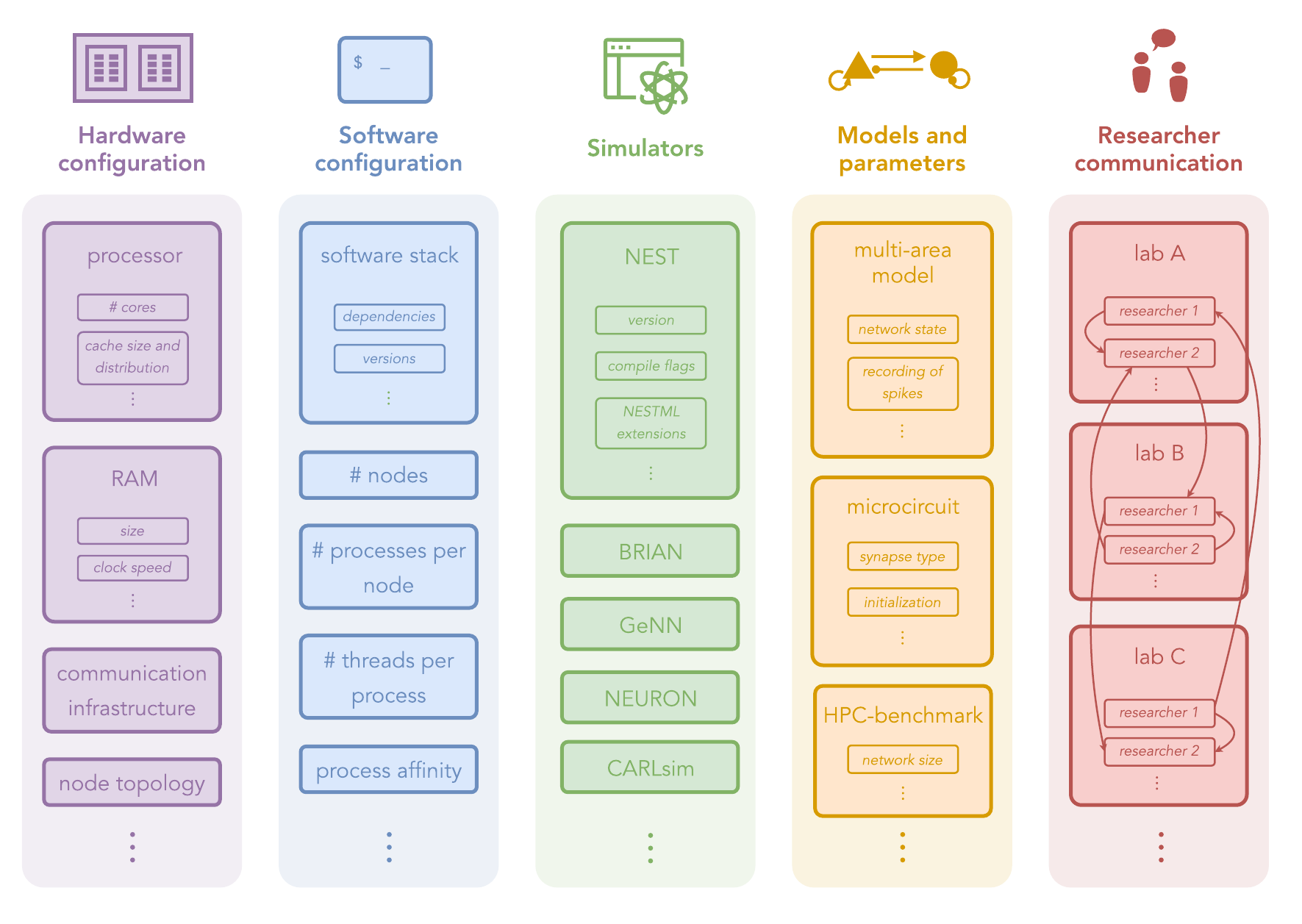}%
\end{center}
\caption{
\textbf{Dimensions of HPC benchmarking experiments with examples from neuronal network simulations.} Hardware configuration: computing architectures and machine specifications. Software configuration: general software environments and instructions for using the hardware. Simulators: specific simulation technologies. Models and parameters: different models and their configurations. Researcher communication: knowledge exchange on running benchmarks.}
\label{fig:problem}
\end{figure}

The omnipresence of benchmarks in studies on simulation technology demonstrates the relevance of efficiency. The intricacy of the benchmarking endeavor, however, not only complicates the comparison between these studies, but also reproducing them. Neuroscientific simulation studies are already difficult to reproduce \citep{Crook13_73,Rougier17_e142,McDougal16_2021,Gutzen18_90,Pauli18,Gleeson19_395}, and benchmarking adds another layer of complexity. Reported benchmarks may differ not only in the structure and dynamics of the employed neuronal network models, but also in the type of scaling experiment, soft- and hardware versions and configurations, as well as in the analysis and presentation of the results. \figref{problem} illustrates the complexity of benchmarking experiments in simulation science and identifies five main dimensions: ``Hardware configuration", ``Software configuration", ``Simulators", ``Models and parameters", and ``Researcher communication".
The following presents examples specific to neuronal network simulations, demonstrating the range of each of the five dimensions.

Different simulators, some with decades of development, allow for large-scale neuroscientific simulations \citep{Brette07_349}. 
We distinguish between simulators that run on conventional HPC systems and those that use dedicated neuromorphic hardware.
Prominent examples of simulators for networks of spiking point-neurons are \texttt{NEST} \citep{Gewaltig_07_11204,Morrison05a,Plesser07_672,Helias12_26,Kunkel2012_5_35,Kunkel14_78,Ippen2017_30,Kunkel2017_11,Jordan18_2,pronold2021routing_efficient_cache} and \texttt{Brian} \citep{Goodman08,Stimberg19_47314} using CPUs;
\texttt{GeNN} \citep{Yavuz16_18854,Knight18_941,Stimberg20_7,Knight21_15,Knight2021_136} and \texttt{NeuronGPU} \citep{Golosio21_627620} using GPUs;
\texttt{CARLsim} \citep{Nageswaran09_791,Richert11_19,Beyeler15_7280424,Chou18_8489326} running on heterogeneous clusters;
and the neuromorphic hardware \texttt{SpiNNaker} \citep{Furber14_652,Rhodes19_20190160}.
\texttt{NEURON} \citep{Carnevale06,Migliore06_119,Lytton2016_2063} and \texttt{Arbor} \citep{Akar19_274} aim for simulating morphologically detailed neuronal networks.

The hardware and software configurations used in published benchmark studies are diverse because both underlie updates and frequent releases. In addition, different laboratories may not have access to the same machines. Therefore, HPC benchmarks are performed on different contemporary compute clusters or supercomputers. For example, \texttt{NEST} benchmarks have been conducted on the systems located at Research Center Jülich in Germany but also on those at the RIKEN Advanced Institute for Computational Science in Japan \cite[e.g., ][]{Helias12_26,Jordan18_2}. To assess the performance of GPU-based simulators, the same simulation is typically run on different GPU devices; from low-end gaming GPUs to those installed in high-end HPC clusters \citep{Knight18_941,Golosio21_627620}. This variety can be beneficial; performing benchmark simulations on only a single system can lead to unwanted optimization towards that type of machine.
However, comparing results across different hard- and software is complicated and requires expert knowledge of the compared technologies in order to draw reasonable conclusions.

The modeling community distinguishes between functional models, where the validation is concerned with the questions if and how well a specific task is solved, and non-functional models, where an analysis of the network structure, dynamics and activity is used for validation. Simulating the same model using different simulation engines often results in activity data which can only be compared on a statistical level. Spiking activity, for example, is typically evaluated based on distributions of quantities such as the average firing rate, rather than on precise spike times \citep{Senk17_243,VanAlbada18_291}. Reasons for that are inevitable differences between simulators such as different algorithms, number resolutions, or random number generators, combined with the fact that neuronal network dynamics is often chaotic, rapidly amplifying minimal deviations  \citep{Sompolinsky88_259,Vreeswijk98_1321,Monteforte10_268104}.
The most frequently used models to demonstrate simulator performance are balanced random networks similar to the one proposed by \cite{Brunel00_183}: generic two-population networks with $80$\% excitatory and $20$\% inhibitory neurons, and synaptic weights chosen such that excitation and inhibition are approximately balanced, similar to what is observed in local cortical networks. Variants differ not only in the parameterization but also in the neuron, synapse and plasticity models, or other details.
Progress in \texttt{NEST} development is traditionally shown by upscaling a model of this type, called ``HPC-benchmark model", which employs leaky integrate-and-fire (LIF) neurons,  alpha-shaped post-synaptic currents, and spike-timing-dependent plasticity (STDP) between excitatory neurons. The detailed model description and parameters can be found in Tables 1--3 of \cite{Jordan18_2}. Other versions include a network of Izhikevich model neurons and STDP \citep{Izhikevich03b} used by \cite{Yavuz16_18854} and \cite{Golosio21_627620}, the COBAHH model with Hodgkin-Huxley type neurons and conductance-based synapses \citep{Brette07_349} used by \cite{Stimberg20_7}, and a version with excitatory LIF and inhibitory Izhikevich model neurons where excitatory synapses are updated with STDP and inhibitory-to-inhibitory connections do not exist is used by \cite{Chou18_8489326}.
Even though balanced random networks are often used for weak-scaling experiments, they describe the anatomical and dynamical features of cortical circuits only at a small spatial scale and the upscaling affects the network dynamics (see \cite{Albada15} as indicated above). At larger scales, the natural connectivity becomes more complex than what is captured by this model type.
Therefore, models of different complexity need to be benchmarked to guarantee that a simulation engine performs well across use cases in the community. In addition to the HPC-benchmark model, this study employs two more elaborate network models: the ``microcircuit model" proposed by \cite{Potjans14_785} and the ``multi-area model" by \cite{Schmidt18_1409}. The microcircuit model is an extension of the balanced random network model with an excitatory and an inhibitory neuron population in each of four cortical layers with detailed connectivity derived from experimental studies. The model spans $\unit[1]{mm^2}$ of cortical surface, represents the cortical layers at their natural neuron and synapse densities, and has recently been used to compare the performance of different simulation engines; for instance, \texttt{NEST} and \texttt{SpiNNaker} \citep{Senk17_243,VanAlbada18_291,Rhodes19_20190160}; \texttt{NEST}, \texttt{SpiNNaker}, and \texttt{GeNN} \citep{Knight18_941}; and \texttt{NEST} and \texttt{NeuronGPU} \citep{Golosio21_627620}. The multi-area model comprises $32$ cortical areas of the visual system where each is represented by an adapted version of the microcircuit model; results are available for \texttt{NEST} \citep{vanAlbada21_47} and \texttt{GeNN} \citep{Knight2021_136}.
Comparing the performance of the same model across different simulators profits from a common model description. The simulator-independent language \texttt{PyNN} \citep{Davison09_10}, for example, enables the use of the same executable model description for different simulator back ends. Testing new technologies only with a single network model is, however, not sufficient for general-purpose simulators and comes with the danger of optimizing the code base for one application, while impairing the performance for others.

Problems to reproduce the simulation outcome or compare results across different studies may not only be technical but also result from a miscommunication between researchers or a lack of documentation. Individual, manual solutions for tracking the hardware and software configuration, the simulator specifics, and the models and parameters used in benchmarking studies have, in our laboratories, proven inefficient when scaling up the number of collaborators. This effect is amplified if multiple  laboratories are involved. Similar inter-dependencies are also present between the other four dimensions of \figref{problem}, making it hard to produce long-term comparable results; the exhibited intricacy of benchmarking is susceptible to errors as, for instance, small details in parameterization or configuration may have a large impact on performance.

Standardizing benchmarks can help to control the complexity but represents a challenge for the fast-moving and interdisciplinary field of computational neuroscience. While the field had some early success in the area of compartmental modeling \citep{Bhalla92a} and \cite{Brette07_349} made initial steps for spiking neuronal networks, neither a widely accepted set of benchmark models nor guidelines for performing benchmark simulations exist. In contrast, benchmarks are routinely employed in computer science, and established metrics help to assess the performance of novel hardware and software. The \texttt{LINPACK} benchmarks \citep{Dongarra03_803}, for example, were initially released in 1979, and the latest version is used to rank the world's top supercomputers by testing their floating-point computing power (TOP500 list). Although this strategy has been successful for many years, it has also been criticized as misguiding hardware vendors towards solutions with high performance in formalized benchmarks but disappointing performance in real-world applications\footnote{\url{https://www.technologyreview.com/2010/11/08/199100/why-chinas-new-supercomputer-is-only-technically-the-worlds-fastest}}. For the closely related field of deep learning, \cite{Dai19_29} summarize seven key properties that benchmarking metrics should fulfill: relevance, representativeness, equity, repeatability, cost-effectiveness, scalability, and transparency. There exist standard benchmarks for machine learning and deep learning applications such as computer vision and natural language processing with standard data sets and a global performance ranking. The most prominent example is \texttt{MLPerf}\footnote{\url{https://mlcommons.org}} \citep{Mattson19_arxiv}. Another example is the High Performance \texttt{LINPACK} for Accelerator Introspection (\texttt{HPL-AI}) benchmark\footnote{\url{https://www.icl.utk.edu/hpl-ai}} which is the mixed-precision counterpart to the \texttt{LINPACK} benchmarks. \cite{Ostrau20_3381772} propose a benchmarking framework for deep spiking neural networks and they compare results obtained with the simulators \texttt{Spikey} \citep{Pfeil13_11}, \texttt{BrainScales} \citep{Schemmel10_1947}, \texttt{SpiNNaker}, \texttt{NEST}, and \texttt{GeNN}.

For measuring and comparing the scaling performance of large-scale neuronal network model simulations, there exists, to our knowledge, no unifying approach, yet. Recently, more laboratories make use of established simulators rather than developing their own, and computing resources have become available and interchangeable. The resulting increase in the size of user-communities comes with the demand for even more flexible and efficient simulators with demonstrated performance. To keep up with this progress, we see the need for a common benchmarking framework. We envision a consistently managed array of standard benchmark models together with standard ways for running them. The five dimensions outlined above lend themselves to a modular framework integrating distinct components which can be updated, extended, or replaced independently. The framework needs to cover all steps of the benchmarking process from configuration, to execution, to handling of results. For enabling comparability and reproducibility, all relevant metadata and data need to be tracked. In this work, we develop a conceptual benchmarking workflow that meets these requirements. For a reference implementation named \texttt{beNNch}, we employ the \texttt{JUBE} Benchmarking Environment\footnote{\label{jube}\url{https://www.fz-juelich.de/ias/jsc/EN/Expertise/Support/Software/JUBE/_node.html}} and the simulator \texttt{NEST} in different versions \citep{Gewaltig_07_11204}, and we assess the time-to-solution for the HPC-benchmark model, the microcircuit model \citep{Potjans14_785}, and the multi-area model \citep{Schmidt18_1409} on the contemporary supercomputer JURECA-DC \citep{Thrnig21_1}. The goal of this study is to set the cornerstone for reliable performance benchmarks facilitating the comparability of results obtained in different settings, and hence, supporting the development of simulators.

The \nameref{sec:results} section of this manuscript formalizes the general concepts of the benchmarking workflow (\secref{workflow_concepts}), implements these concepts into a reference benchmarking framework for the \texttt{NEST} simulator (\secref{reference_implementation}), and applies the framework to generate and compare benchmarking data, thereby making a case for the relevance of benchmarking for simulator development (\secref{use_cases}).
After a discussion of our results in \secref{discussion}, \secref{materials_and_methods} provides details of specific performance optimizations addressed in this work.

\newpage

\section{Results}
\label{sec:results}
\subsection{Workflow concepts}
\label{sec:workflow_concepts}
\begin{figure}[h!]
\begin{center}
\includegraphics{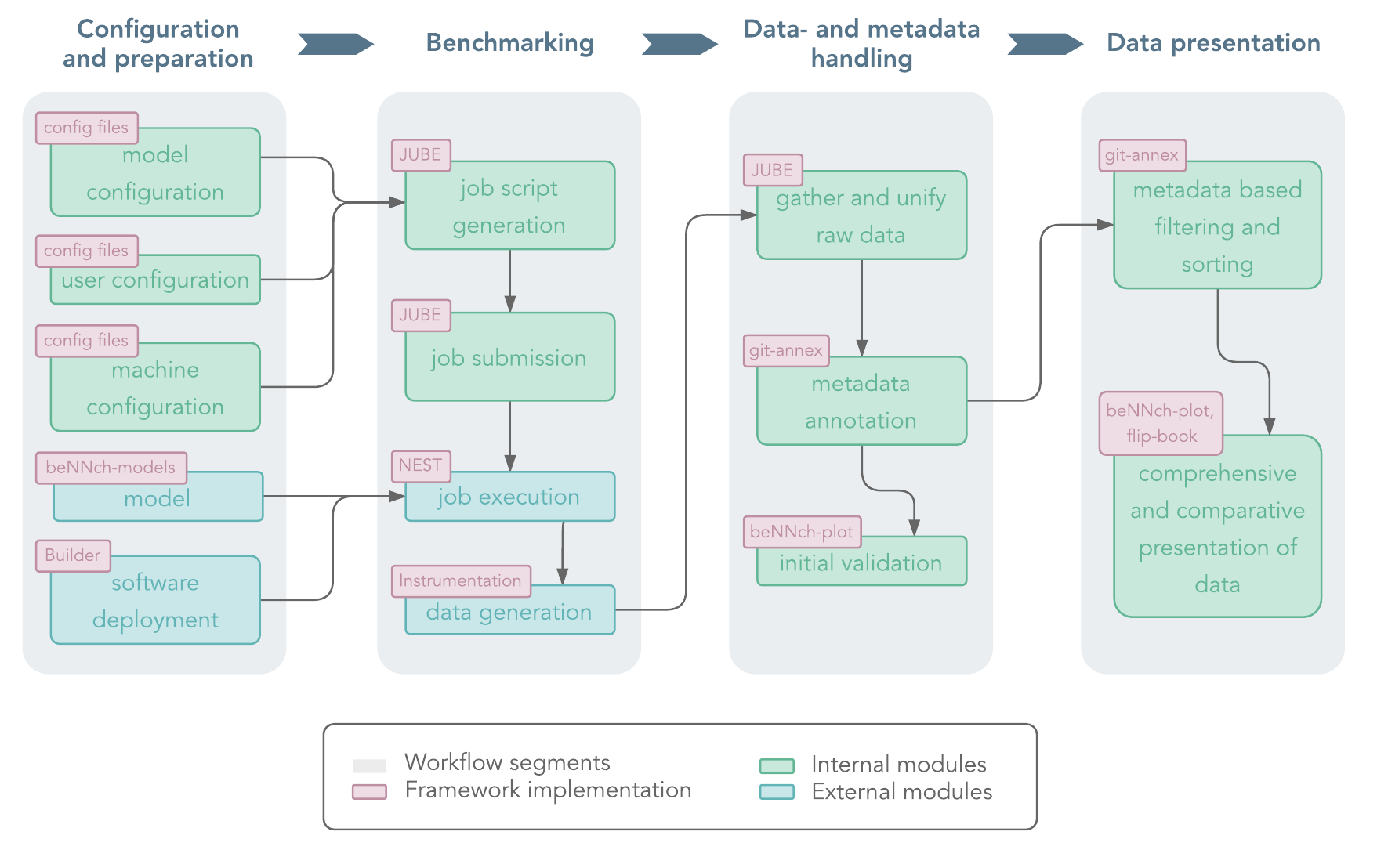}%
\end{center}
\caption{
\textbf{Conceptual overview of the proposed benchmarking workflow.}
Light gray boxes divide the workflow into four distinct segments, each consisting of multiple modules. Internal modules are shown in mint and external ones in cyan. Pink boxes indicate their respective realization in our reference implementation.
}
\label{fig:concepts}
\end{figure}

We devise a generic workflow for performance benchmarking applicable to simulations running on conventional HPC architectures. The conceptual workflow depicted in \figref{concepts} consists of four segments which depend on each other in a sequential fashion. The segments are subdivided into different modules which are related to the specific realizations used in our reference implementation of the workflow (\secref{reference_implementation}). We use the term ``workflow" to describe abstract concepts that are of general applicability with regard to benchmarking efforts, and ``framework" to refer to the concrete software implementation we have developed.
Further, we make the distinction between ``internal" and ``external" modules. Internal modules are considered essential building blocks of the workflow while external modules can be exchanged more readily.
The following introduces each of the workflow's conceptual segments and explains how the proposed solution addresses the identified problems (cf. \figref{problem}). 

\subsubsection{Configuration and preparation}
\label{sec:conf_and_prep}

The first of the four workflow segments consists of five distinct modules that together set up all necessary prerequisites for the simulation. First, the installation of the simulation software and its dependencies is handled by ``software deployment", while ``machine configuration" specifies parameters that control the simulation experiment conditions, as for example, how many compute nodes to reserve.  Together, these two modules target the problem dimensions ``hardware configuration", ``software configuration", and ``simulators". Addressing ``models and parameters", the module ``model" provides the network model implementation, while ``model configuration" allows for passing parameters to the model such as the biological model time to be simulated, thereby separating the model from its parameters. Finally, the ``user configuration" module confines user-specific data, such as file paths or compute budgets, to a single location.

\subsubsection{Benchmarking}
\label{sec:benchmarking}

The second segment encompasses all modules related to actually running the benchmark simulation. Compute clusters typically manage the workload of the machine via queuing systems; therefore, compute-intensive calculations are submitted as jobs via scripts which define resource usage and hold instructions for carrying out the simulation. In the workflow, this is handled by the module aptly named ``job script generation". Here, the first link between modules comes into play: the workflow channels model, user and machine configuration to create a job script and subsequently submit the script to the job queue via the module ``job submission". With the simulator software prepared by the software-deployment module, ``job execution" performs the model simulation given the job-submission parameters. While a simulation for neuroscientific research purposes would at this point focus on the output of the simulation, for example, neuronal spike times or voltage traces, benchmarking is concerned with the performance results. These are recorded in the final benchmarking module called ``data generation".

\subsubsection{Data- and metadata handling}
\label{sec:data_metadata}

A core challenge in conducting performance benchmarks is the handling of all produced data and metadata. While the former type of data here refers to the results of the performance measurements, the latter is an umbrella term describing the circumstances under which the data was recorded according to the dimensions of benchmarking (\figref{problem}). Since executing multiple simulations using different configurations, software, hardware, and models is an integral part of benchmarking, data naturally accumulates. Recording the variations across these dimensions leads to a multitude of metadata that needs to be associated to the measured data. Standardized formats for both types of data make the results comparable for researchers working with the same underlying simulation technology. The workflow segment ``Data- and metadata handling" proposes the following solution. First, the raw performance data, typically stemming from different units of the HPC system, are gathered and unified into a standardized format, while the corresponding metadata is automatically recorded. Next, the metadata is associated to the unified data files, alleviating the need for manually keeping track of parameters, experiment choices and software environment conditions. While there are different possible solution for this, attaching the relevant metadata directly to the performance-data files simplifies filtering and sorting of results.
Finally, ``initial validation" allows for a quick glance at the results such that erroneous benchmarks can be swiftly identified.

\subsubsection{Data presentation}
\label{sec:data_presentation}

This final workflow segment addresses the challenge of making the benchmarking results accessible and comparable such that meaningful conclusions can be drawn, thereby aiming to cope with the complexity that ``Researcher communication" introduces.
In a first step, ``metadata based filtering and sorting" allows the user to dynamically choose the results to be included in the comparison.
Here, dynamic means that arbitrary cuts through the hypercube of metadata dimensions can be selected such that the filtered results only differ in metadata fields of interest. Second, the data is presented in a format for which switching between benchmarks is intuitive, key metadata is given alongside the results, and data representation is standardized. The presentation of data should be comprehensive, consistent, and comparative such that the benchmarking results are usable in the long term. Thereby, the risk of wasting resources through re-generation of results is eliminated, making the corresponding software development more sustainable.
\subsection{beNNch: A reference implementation}
\label{sec:reference_implementation}
Building on the fundamental workflow concepts developed in \secref{workflow_concepts}, we introduce a reference implementation for modern computational neuroscience: \texttt{beNNch}\footnote{\url{https://github.com/INM-6/beNNch}}\textemdash a benchmarking framework for neuronal network simulations.
The framework serves not only as a proof-of-concept, but also provides a software tool that can be readily used by neuroscientists and simulator developers.
While \texttt{beNNch} is built such that plug-ins for any neuronal network simulator can be developed, we specifically implement compatibility with the \texttt{NEST} simulator \citep{Gewaltig_07_11204} designed for simulating large-scale spiking neuronal network models.
In the following subsections, we detail software tools, templates, technologies, and user specifications needed to apply \texttt{beNNch} for benchmarking \texttt{NEST} simulations.
Each of the conceptual modules of \figref{concepts} is here associated with a concrete reference.

\subsubsection{Builder}
\label{sec:builder}

Reproducible software deployment is necessary for repeatability and comparability of the benchmarks. In favor of the usability of the benchmarking framework, however, we need to abstract non-relevant information on the hardware architecture and the software tool chain. The tool set is required to install software in a platform independent way and should not depend on a particular flavor of the operating system, the machine architecture or overly specific software dependencies. Additionally, it needs to be able to make use of system-provided tools and libraries, for example, to leverage machine specific MPI implementations.
\texttt{beNNch} uses the tool \texttt{Builder}\footnote{\url{https://github.com/INM-6/Builder}} for this purpose. Given a fixed software stack and hardware architecture, \texttt{Builder} provides identical executables by deriving the install instructions from ``plan files". Integration with other package management systems such as \texttt{easy\_build} \citep{Geimer14_easy} or \texttt{Spack} \citep{Gamblin15_spack} is achieved by using the same environment module systems\footnote{\url{https://modules.readthedocs.io} and \url{http://lmod.readthedocs.io}}. Thereby, the required user interaction is minimized and, from a user perspective, installation reduces to the configuration of installation parameters. Given a specified variation of the software to be benchmarked, \texttt{beNNch} calls \texttt{Builder} to deploy the requested software. In doing so, \texttt{Builder} checks whether the software is already available and otherwise installs it according to the specifications in the plan file. The depth to which required dependencies need to be installed and which mechanisms are used depend on the conventions and prerequisites available at each execution site. For any installation, the used software stack\textemdash including library versions, compiler versions, compile flags, etc.\textemdash are recorded as metadata.

\subsubsection{NEST}
\label{sec:NEST}

\texttt{beNNch} implements compatibility with the \texttt{NEST} simulator \citep{Gewaltig_07_11204}, enabling the performance benchmarking of neuronal network simulations at the resolution of single neurons.
The \texttt{NEST} software is complex, and the consequences of code modifications for performance are often hard to predict.
\texttt{NEST} has an efficient C++ kernel, but network models and simulation experiments are defined via the user-friendly Python interface \texttt{PyNEST} \citep{Eppler09_12,Zaytsev14_23}.
To parallelize simulations, \texttt{NEST} provides two methods: for distributed computing, \texttt{NEST} employs the Message Passing Interface \citep[MPI, ][]{MPIForum09}, and for thread-parallel simulation, \texttt{NEST} uses OpenMP \citep{OpenMPSpec}.

\subsubsection{Instrumentation}
\label{sec:instrumentation}

\begin{figure}[t]
\begin{center}
\includegraphics{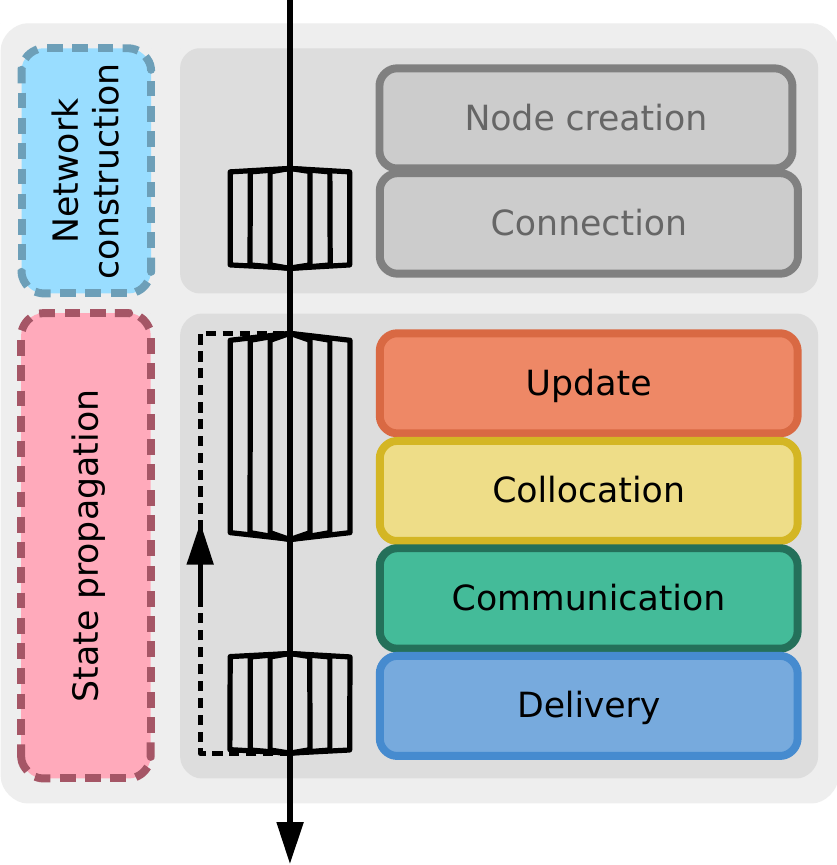}%
\end{center}
\caption{
\textbf{Instrumentation to measure time-to-solution.}
Successive phases of a \texttt{NEST} simulation; time is indicated by top-down arrow.
Fanning arrows denote parallel operation of multiple threads.
The main phases network construction (cyan) and state propagation (pink) are captured by external timers on the Python level.
Built-in \texttt{NEST} timers on the C++ level measure sub-phases: node creation and connection (both grey, not used in benchmark plots); update (orange), collocation (yellow), communication (green), and delivery (blue).
The sub-phases of the state propagation are repeated until the simulation is finished as shown by the dashed arrow connecting delivery and update.
}
\label{fig:timers}
\end{figure}

We focus our performance measurements on the time-to-solution.
Acquiring accurate data on time consumption is critical for profiling and benchmarking.
To this end, we make use of two types of timers to collect this data:
The timers are either built-in to \texttt{NEST} on the C++ level, or they are included on the Python level as part of the \texttt{PyNEST} network-model description.
The latter type of timers are realized with explicit calls to the function \texttt{time.time()} of the Python Standard Library's  \texttt{time}.
To achieve consistency throughout the framework, we use standardized variable names for the different phases of the simulation.
\figref{timers} shows the simulation flow of a typical \texttt{NEST} simulation.
During ``network construction", neurons and auxiliary devices for stimulation and recording are created and subsequently connected according to the network-model description.
Afterwards, in the course of ``state propagation", the network state is propagated in a globally time-driven manner.
This comprises four main phases which are repeated until the entire model time has been simulated: update of neuronal states, collocation of spikes in MPI-communication buffers, communication of spikes, and delivery of the received spikes to their respective thread-local targets.
\texttt{NEST}'s built-in timers provide a detailed look into the contribution of all four phases of state propagation, while timers on the Python level measure network construction and state propagation.

In \texttt{NEST}, the postsynaptic connection infrastructure is established during the ``connection" phase.
However, the presynaptic counterpart is typically only set up at the beginning of the state propagation phase \citep[see][for details]{Jordan18_2}.
In this work, we trigger this step deliberately and include it in our measurement of network-construction time rather than state-propagation time.
Besides, it is common practice to introduce a short pre-simulation before the actual simulation to give the network dynamics time to level out; the state propagation phase is only recorded when potential startup transients have decayed  \citep{Rhodes19_20190160}. The model time for pre-simulation can be configured via a parameter in \texttt{beNNch}.
For simplicity, \figref{timers} does not show this pre-simulation phase.

\subsubsection{beNNch-models}
\label{sec:beNNch-models}

We instantiate the ``model" module with the repository \texttt{beNNch-models}\footnote{\url{https://github.com/INM-6/beNNch-models}} which contains a collection of \texttt{PyNEST} neuronal network models, i.e., models that can be simulated using the Python interface of \texttt{NEST} \citep{Eppler09_12}.
In principle, any such model can be used in conjunction with \texttt{beNNch}; only a few adaptations are required concerning the interfacing.
On the input side, the framework needs to be able to set relevant model parameters.
For recording the performance data, the required Python timers (\secref{instrumentation}) must be incorporated.
On the output side, the model description is required to include instructions to store the recorded performance data and metadata in a standardized format.
Finally, if a network model is benchmarked with different \texttt{NEST} versions that require different syntax, the model description also needs to be adjusted accordingly.
Instructions on how to adapt existing models are provided in the documentation of \texttt{beNNch-models}.

The current version of \texttt{beNNch} provides benchmark versions of three widely studied spiking neuronal network models:
The two-population HPC-benchmark model\footnote{original repository: \url{https://github.com/nest/nest-simulator/blob/master/pynest/examples/hpc_benchmark.py}}, the microcircuit model\footnote{original repository: \url{https://github.com/nest/nest-simulator/tree/master/examples/nest/Potjans_2014}} by \cite{Potjans14_785} representing $\unit[1]{mm^2}$ of cortical surface with realistic neuron and synapse densities, and the multi-area model\footnote{original repository: \url{https://github.com/INM-6/multi-area-model}} by \cite{Schmidt18_1409,Schmidt18_e1006359} consisting of $32$ microcircuit-like interconnected networks representing different areas of visual cortex of macaque monkey.
The model versions used for this study employ the required modifications described above.

\subsubsection{config files}
\label{sec:config_files}

When executing benchmarks, the main user interaction with \texttt{beNNch} consists of defining the characteristic parameters.
We separate this from the executable code by providing \texttt{yaml}-based templates for ``config files"  to be customized by the user.
Thereby, the information that defines a benchmark experiment is kept short and well arranged, limiting the number of files a user needs to touch and reducing the risk of user errors on the input side.
\lstref{config_file_example} presents an excerpt from such a config file which has distinct sections to specify model, machine, and software parameters.
While some parameters are model specific, standardized variable names are defined for parameters that are shared between models.

\begin{algorithm}[t]
\lstset{language=myYAML}
\begin{lstlisting}
parameterset:

    - name: model_parameters
      parameter:
       # can be either "metastable" or "ground"
       - {name: network_state, type: string, _: "metastable"}
       # biological model time to be simulated in ms
       - {name: model_time_sim, type: float, _: "10000."}
       # "weak" or "strong" scaling
       - {name: scaling_type, _: "strong"}
    
    - name: machine_parameters
      parameter:
       # number of compute nodes
       - {name: num_nodes, type: int, _: "4,8,12,16,24,32"}  
       # number of MPI tasks per node
       - {name: tasks_per_node, type: int, _: "8"}
       # number of OpenMP threads per task
       - {name: threads_per_task, type: int, _: "16"}
       
    - name: software_parameters
      parameter:
       # simulator used for executing benchmarks
       - {name: simulator, _: "nest-simulator"}  
       # simulator version
       - {name: version, _: "3.0"}
\end{lstlisting}
\caption{
\textbf{Excerpt of a config file in \texttt{yaml}-format for setting model, machine, and software parameters for benchmarking the multi-area model.}
When giving a list (e.g., for \texttt{num\_nodes}), a job for each entry of the list is created.
\textbf{Model parameters}: \texttt{network\_state} describes particular model choices that induce different dynamical fixed points; \texttt{model\_time\_sim} defines the total model simulation time in $\unit[]{ms}$; \texttt{scaling\_type} sets up the simulation for either a weak- or a strong-scaling experiment. The former scales the number of neurons linearly with the used resources which might be ill-defined for anatomically constrained models.
\textbf{Machine parameters}: \texttt{num\_nodes} defines the number of nodes over which the scaling experiment shall be performed; \texttt{tasks\_per\_node} and \texttt{threads\_per\_task} specify the number of MPI tasks per node and threads per MPI task respectively.
\textbf{Software parameters}: \texttt{simulator} and \texttt{version} describe which version of which simulator to use (and to install if not yet available on the machine).
}
\label{lst:config_file_example}
\end{algorithm}

\subsubsection{JUBE}
\label{sec:jube}

At this point, the first segment of the benchmarking workflow (\figref{concepts}) is complete and hence all necessary requirements are set up: the software deployment provides the underlying simulator (here: \texttt{NEST} with built-in instrumentation), the models define the simulation, and the configuration specifies the benchmark parameters.
This information is now processed by the core element of the framework: generating and submitting simulation jobs and gathering and unifying the obtained performance data.
We construct this component of \texttt{beNNch} around the \texttt{xml}-based \texttt{JUBE}\footnoteref{jube} software tool using its \texttt{yaml} interface.
Built around the idea of benchmarking, \texttt{JUBE} can fulfill the role of creating job scripts from the experiment, user and machine configuration, their subsequent submission, as well as gathering and unifying of the raw data output.
Here, we focus on the prevalent scheduling software \texttt{SLURM} \citep{Yoo03_44}, but extensions to allow for other workload managers would be straightforward to implement.
Our approach aims at high code re-usability. Model specific code is kept to a minimum, and where necessary, written in a similar way across models.
Adhering to a common interface between \texttt{JUBE} scripts and models facilitates the integration of new models, starting from existing ones as a reference.
Since \texttt{JUBE} can execute arbitrary code, we use it to also record metadata in conjunction with each simulation.
This includes specifications of the hardware architecture as well as parameters detailing the run and model configuration.

\subsubsection{git-annex}
\label{sec:git-annex}

Without a mature strategy for sharing benchmark results, communication can be a major obstacle. Typically, each researcher has their preferred workflow, thus results are shared over different means of communication, for example, via email attachments, cloud-based storage options, or \texttt{git} repositories.
This makes it difficult to maintain an overview of all results, especially if researchers from different labs are involved.
Ideally, results would be stored in a decentralized fashion that allows for tracking the history of files while allowing on-demand access for collaborators.
To this end, we use \texttt{git-annex}\footnote{\url{https://git-annex.branchable.com}} as a versatile base technology; it synchronizes file information in a standard \texttt{git} repository while keeping the content of large files in a separate object store, thereby keeping the repository size at a minimum.
\texttt{git-annex} is supported by the GIN platform\footnote{\url{https://gin.g-node.org}} which we employ for organizing our benchmark results.
In addition, it allows for metadata annotation: instead of relying on separate files that store the metadata, \texttt{git-annex} can directly attach them to the data files, thereby implementing the ``metadata annotation" module.
Previously this needed to be cataloged by hand, whereas now the framework allows for an automatic annotation, reducing the workload on researchers and thus probability of human mistakes.

A difficult task when scaling up the usage of the framework and, by extension, handling large amounts of results, is providing an efficient way of dealing with user queries for specific benchmark results.
Attaching the metadata directly to the performance data not only reduces the visible complexity of the repository, but also provides an efficient solution: \texttt{git-annex} implements a native way of selecting values for metadata keys via \texttt{git-annex} ``views", automatically and flexibly reordering the results in folders and sub-folders accordingly. For example, consider the case of a user specifying the \texttt{NEST} version to be \texttt{3.0}, the \texttt{tasks\_per\_node} to be either $4$ or $8$, and the \texttt{network\_state} to be either \texttt{metastable} or \texttt{ground}. First, \texttt{git-annex} filters out metadata keys for which only a single value is given; in our example, only benchmarks conducted with \texttt{NEST} version \texttt{3.0} remain. Second, a hierarchy of directories is constructed with a level for each metadata key for which multiple options are given. Here, the top level contains the folders ``$4$" and ``$8$", each containing sub-folders \texttt{metastable} and \texttt{ground} where the corresponding results reside.
However, it may be difficult to judge exactly what metadata is important to collect; oftentimes, it is only visible in hindsight that certain metadata is relevant for the simulation performance.
Therefore, recording as much metadata as possible would be ideal, allowing for retrospective investigations if certain metadata becomes relevant after run time.
Importantly, a balance needs to be struck between recording large amounts of metadata and keeping the volume of annotations manageable.
In our implementation, we choose to solve this issue by recording detailed metadata about the system, software, and benchmarks, but only attaching what we currently deem relevant for performance to the data.
The remaining metadata is archived and stored alongside the data, thereby sacrificing ease of availability for a compact format.
This way, if future studies discover that a certain hardware feature or software parameter is indeed relevant for performance, the information remains accessible also for previously simulated benchmarks while staying relatively hidden otherwise.
Furthermore, using \texttt{git} as a base technology allows to collect data sets provided by different researchers in a curated fashion by using well established mechanisms like branches and merge-request reviews.
This use of \texttt{git-annex} thereby implements the ``metadata based filtering and sorting" module of \figref{concepts}.

\subsubsection{beNNch-plot}
\label{sec:beNNch-plot}

To enable a comparison between plots of benchmark results across the dimensions illustrated in \figref{problem} it is paramount to use the same plotting style. 
To this end, we have developed the standalone plotting package \texttt{beNNch-plot}\footnote{\url{https://github.com/INM-6/beNNch-plot}} based on \texttt{matplotlib} \citep{Hunter07}.
Here, we define a set of tools to create individual plot styles that can be combined flexibly by the user.
The standardized definitions of performance measures employed by \texttt{beNNch} directly plug into this package.
In addition, \texttt{beNNch-plot} includes default plot styles that can be readily used, and provides a platform for creating and sharing new ones.
\texttt{beNNch} utilizes the default plot styles of \texttt{beNNch-plot} for both initial validation\textemdash a preliminary plot offering a quick glance at the results, thereby enabling a swift judgement whether any problems occurred during simulation\textemdash and visualization of the final results.

\subsubsection{flip-book}
\label{sec:flip-book}

When devising a method of presenting benchmark results we found the following aspects to be of crucial relevance for our purposes.
First, it should be possible to navigate the results such that plots are always at the same screen position and have the same dimensions, thereby minimizing the effort to visually compare results.
To achieve such a format, we decided to create a flip-book in which each slide presents the results of one experiment.
Second, relevant metadata should be displayed right next to the plots. This can include similarities across the runs, but more importantly should highlight the differences.
As each user might be interested in different comparisons, we let the user decide which kind of metadata should be shown.
Third, it should be easy to select only the benchmarks of interest in order to keep the number of plots small.
This is already handled by the filtering provided by \texttt{git-annex} views as described in \secref{git-annex}.
As an underlying technology for programmatically creating \texttt{HTML} slides we use \texttt{jupyter notebooks}\footnote{\url{https://jupyter.org}} in conjunction with the open source \texttt{HTML} presentation framework \texttt{reveal.js}\footnote{\url{https://github.com/hakimel/reveal.js}}. An exemplary flip-book containing the \texttt{NEST} performance results described in this work is published alongside the \texttt{beNNch} repository\footnote{\label{flip-book_url}\url{https://inm-6.github.io/beNNch}}.
By respecting these considerations, our proposed solution offers a way of sharing benchmarking insights between researchers that is both scalable and flexible.

\subsection{Use case: NEST development}
\label{sec:use_cases}
This section illustrates the relevance of performance benchmarks for the development of neuronal network simulators with the example of recent changes to the \texttt{NEST} code base. The series of \texttt{NEST 2.X} releases includes enhancements, bug fixes, and contributions to maintenance with only marginal effects on the \texttt{PyNEST} user interface \citep{Eppler09_12}. Performance-related updates to the simulation kernel are accomplished under the hood.
The 3g kernel \citep{Helias12_26,Kunkel2012_5_35} is in use from \texttt{NEST 2.2.0} \citep{Nest22}. \texttt{NEST 2.12.0} \citep{Nest2120} introduces the 4g kernel \citep{Kunkel14_78} which implements novel data structures allowing for an efficient and flexible representation of sparse network connectivity on highly distributed computing systems such as supercomputers. The 5g kernel \citep{Jordan18_2} in \texttt{NEST 2.16.0} \citep{Nest2160} continues this direction of development toward an optimal usage of HPC systems for large-scale simulations by disentangling the memory usage per compute node from the total network size. The transition from \texttt{NEST 2} to \texttt{NEST 3} corresponds to a refurbishment of the simulator code which also breaks the backwards compatibility of the user interface. While improved high-level functionality and parameter handling are the primary goals of this transition, the 5g kernel is supposed to remain. In the past, performance changes due to kernel updates have been predominantly assessed using the HPC-benchmark model.
The performance of the \texttt{NEST 3.0} release candidate (``\texttt{3.0rc}"), however, is in addition evaluated with the microcircuit and multi-area model which exhibit a more complex connectivity structure and a different distribution of synaptic delays. In this way, so far undetected performance bottlenecks are discovered and subsequently resolved, leading to the official release \texttt{NEST 3.0} \citep{Nest30}.

\begin{table}[t]
    \renewcommand\arraystretch{1.75}
    \centering
    \begin{tabularx}{\textwidth}{@{\hspace{20pt}}l@{\hspace{20pt}}@{\hspace{20pt}}X@{\hspace{20pt}}} 
         Shorthand notation of \texttt{NEST} version & Description \\
         \toprule
         \texttt{2.20.2} & Official \texttt{2.20.2} release \citep{Nest2202}\\ 
         \midrule
         \texttt{3.0rc} & Release candidate for \texttt{3.0}\\ 
         \midrule
         \texttt{3.0rc+ShrinkBuff} & \texttt{3.0rc} plus shrinking MPI buffers \\ 
         \midrule
         \texttt{3.0rc+ShrinkBuff+SpikeComp} & \texttt{3.0rc+ShrinkBuff} plus spike compression \\ 
         \midrule
         \texttt{3.0} & Official \texttt{3.0} release \citep{Nest30} $=$ \texttt{3.0rc+ShrinkBuff+SpikeComp} plus neuronal input buffers with multiple channels \\
    \end{tabularx}
    \caption{Shorthand notation and description of \texttt{NEST} versions used in this work.}
    \renewcommand\arraystretch{1.}
    \label{tab:nest_version_names}
\end{table}

\begin{figure}[t]
\begin{center}
\includegraphics{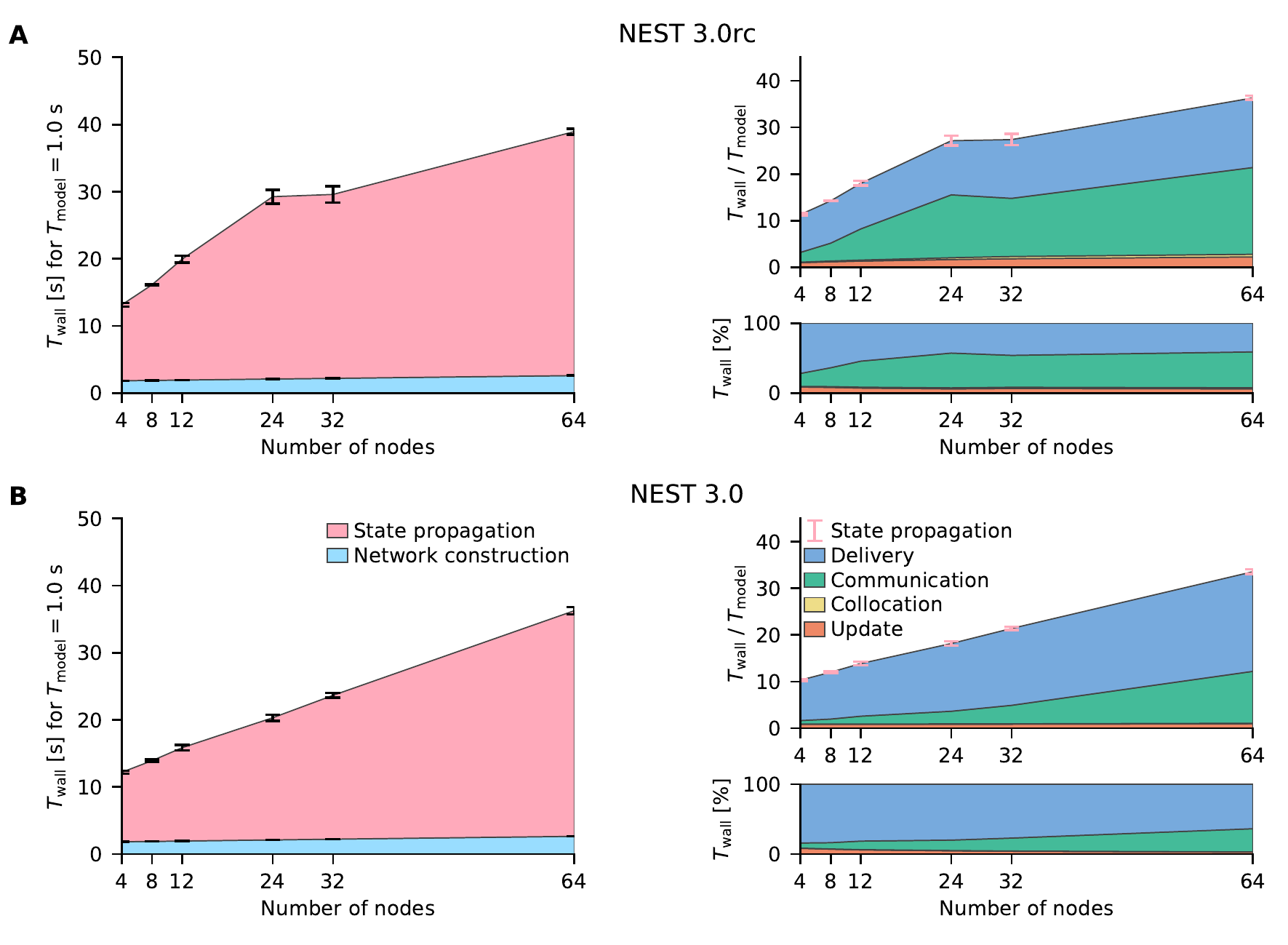}%
\end{center}
\caption{
\textbf{Weak-scaling performance of the HPC-benchmark model on JURECA-DC.}
\textbf{A}~\texttt{NEST~3.0rc}.
The left graph shows the absolute wall-clock time $T_\mathrm{wall}$ measured with Python-level timers for both network construction and state propagation (legend in panel \textbf{B}); the model time is $T_\mathrm{model}=\unit[1]{s}$. Error bars indicate variability across three simulation repeats with different random seeds.
The top right graph displays the real-time factor defined as wall-clock time normalized by the model time. Built-in timers resolve four different phases of the state propagation (legend in panel \textbf{B}): update, collocation, communication, and delivery. Pink error bars show the same variability of state propagation as the left graph. The lower right graph shows the relative contribution of these phases to the state-propagation time. Same colors used for phases as in \figref{timers}.
\textbf{B}~\texttt{NEST~3.0}. Same display as panel A.
}
\label{fig:hpc-benchmark}
\end{figure}

Here, we use \texttt{beNNch} to outline crucial steps of the development from the release candidate \texttt{NEST~3.0rc} to the final \texttt{NEST 3.0} and also discuss improvements compared to the latest \texttt{NEST 2} version \citep[\texttt{NEST 2.20.2}, ][]{Nest2202}. \tabref{nest_version_names} summarizes the \texttt{NEST} versions employed in this study. Our starting point is the weak-scaling experiments of the HPC-benchmark model \citep{Jordan18_2}; the times for network construction and state propagation as well as the memory usage remain almost constant with the newly introduced 5g kernel (see their Figures 7 and 8). \figref{hpc-benchmark} shows similar benchmarks of the same network model conducted with \texttt{beNNch} using the release candidate in \figref{hpc-benchmark}A and the final release in \figref{hpc-benchmark}B.
The graph design used here corresponds to the one used in the flip-book format by the framework. A flip-book version of the results shown in this work can be accessed via the GitHub Pages instance of the \texttt{beNNch} repository\footnoteref{flip-book_url}.
While the release candidate in \figref{hpc-benchmark}A exhibits growing state-propagation times when increasing the number of nodes, network-construction times stay constant and are, for $T_{\mathrm{model}} = \unit[1]{s}$, small, making up less than $10\%$ of the total simulation time. The phases ``delivery" and ``communication" both contribute significantly to the state-propagation time. \cite{Jordan18_2} report real-time factors of about $500$ (e.g., their Figure 7C) in contrast to values smaller than $40$ shown here and their simulations are by far dominated by the delivery phase (see their Figure 12). A comparison of our data and the data of \cite{Jordan18_2} is not straightforward due to the inherent complexity of benchmarking and we will here emphasize a few concurring aspects: First, \cite{Jordan18_2} run their benchmarks on the dedicated supercomputers JUQUEEN \citep{stephan2015juqueen} and K Computer \citep{Miyazaki12} while our benchmarks use the recent cluster JURECA-DC \citep{Thrnig21_1}. Each compute node of the BlueGene/Q system JUQUEEN is equipped with a $16$-core IBM PowerPC A2 processor running at $\unit[1.6]{GHz}$ and each node of the K Computer has an $8$-core SPARC64 VIIIfx processor operating at $\unit[2]{GHz}$; both systems provide $\unit[16]{GB}$ RAM per node. In contrast, the JURECA-DC cluster employs compute nodes consisting of two sockets, each housing a 64-core AMD EPYC Rome $7742$ processors clocked at $\unit[2.2]{GHz}$, that are equipped with $\unit[512]{GB}$ of DDR4 RAM. Here, nodes are connected via an InfiniBand HDR100/HDR network. Second, \cite{Jordan18_2} use $1$ MPI process per node and $8$ threads per process while our simulations are performed throughout this study with $8$ MPI processes per node and $16$ threads per process. Third, \cite{Jordan18_2} simulate $18,000$ neurons per MPI process while we only simulate $11,250$ neurons per process. This list of differences is not complete and only aims to illustrate that potential discrepancies in benchmarking results may be explained by differences in hardware, software, simulation and model configuration, and other aspects exemplified in \figref{problem}.

\begin{figure}[h!]
\begin{center}
\includegraphics{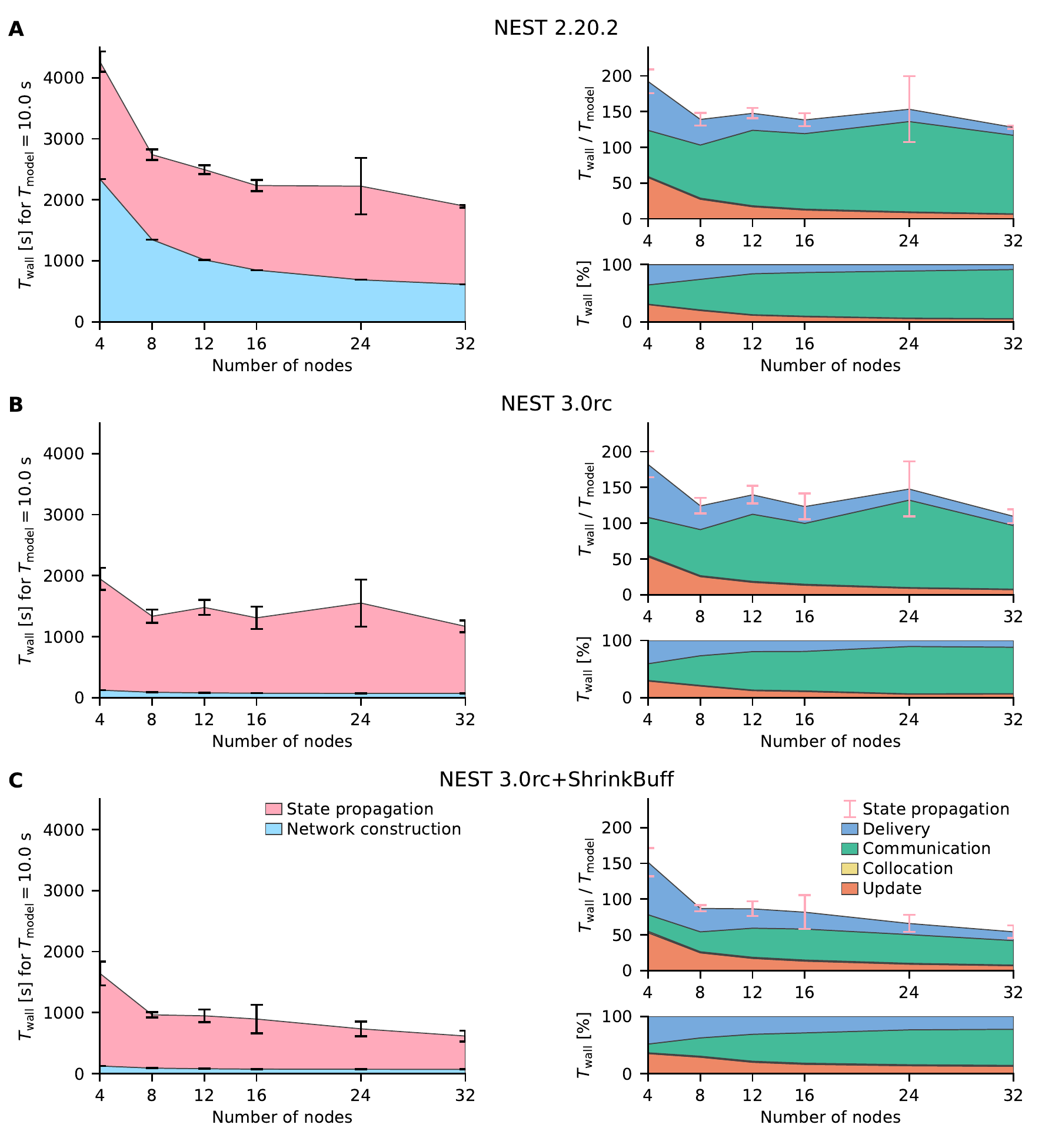}
\end{center}
\caption{
\textbf{Strong-scaling performance of the multi-area model on JURECA-DC.} Same display as in \figref{hpc-benchmark}. The multi-area model is simulated in its meta-stable state leading to a high amount of spikes that are communicated. The model time is $T_\mathrm{model}=\unit[10]{s}$. Simulations are repeated for ten different random seeds. \textbf{A}~\texttt{NEST~2.20.2} (latest \texttt{NEST~2} release). \textbf{B}~\texttt{NEST~3.0} release candidate. \textbf{C}~\texttt{NEST~3.0} release candidate with shrinking MPI buffers.}
\label{fig:MAM}
\end{figure}

Having demonstrated that \texttt{beNNch} can perform weak-scaling experiments of the HPC-benchmark model as done in previous publications, we next turn to strong-scaling benchmarks of the multi-area model \citep{Schmidt18_1409}. To fulfill the memory requirements of the model, at least three compute nodes of JURECA-DC are needed; here, we choose to demonstrate the scaling on four to $32$ nodes. Initially, we compare the latest \texttt{NEST~2} version (\figref{MAM}A) with the release candidate for \texttt{NEST~3.0} (\figref{MAM}B). The improved parameter handling implemented in \texttt{NEST~3.0rc} reduces the network-construction time. However, the communication phase here largely dominates state propagation in both \texttt{NEST} versions shown; both use the original 5g kernel.
Previous simulations of the HPC-benchmark model have not identified the communication phase as a bottleneck \cite[][Figure 12]{Jordan18_2}. Communication only becomes an issue when then smallest delay in the network is of the same order as the computation step size because \texttt{NEST} uses the smallest delay as the communication interval for MPI. While the HPC-benchmark model uses $\unit[1.5]{ms}$ for all connections\textemdash which is a good estimate for inter-area connections\textemdash the multi-area model and microcircuit use distributed delays with a lower bound of $\unit[0.1]{ms}$ leading to a fifteen-fold increase in the number MPI communication steps.

The following identifies and eliminates the main cause for the large communication time in case of the multi-area model, thus introducing the first out of three performance-improving developments applied to \texttt{NEST~3.0rc}. Cross-node communication, handled in \texttt{NEST} by MPI, needs to strike a balance between the amount of messages to transfer and the size of each message. The size of the MPI buffer limits the amount of data that fits into a single message, and is therefore the main parameter controlling this balance.
Ideally, each buffer would fit exactly the right amount of information by storing all spikes of the process relevant for the respective communication step. Due to overhead attached to operating on additional vectors, a scheme in which the buffer size adapts precisely for each MPI process for each communication step can be highly inefficient. Therefore, in cases where communication is roughly homogeneous, it is advantageous to keep the exchanged buffer between all processes the same size, as is implemented in \texttt{NEST~3.0rc}. While buffer sizes are constant across processes, \texttt{NEST} does adapt them over time to minimize the number of MPI communications. Concretely, whenever the spike information that a process needs to send exceeds what fits into one buffer, the buffer size for the next communication step is increased. However, the original 5g kernel of \texttt{NEST} does not shrink buffer sizes. In networks such as the multi-area model, the firing is not stationary over time; transients of high activity propagate through the network \citep{Schmidt18_1409}. In general, startup transients may cause high spike rates only in the beginning of a simulation unless the network is carefully initialized \citep{Rhodes19_20190160}. If the rates decrease, the spiking information becomes much smaller than the available space in the MPI buffer. Consequently, the original 5g kernel preserves unnecessarily large buffer sizes which results in the communication of useless data. To address this issue, a mechanism for automatically  shrinking the buffer sizes has been introduced. For details see \secref{methods_adaptive-buffers} ``\nameref{sec:methods_adaptive-buffers}". The release candidate with the implementation of shrinking MPI buffers (\texttt{NEST~3.0rc+ShrinkBuff}) approximately halves the time spent in the communication phase compared to the original implementation (compare \figref{MAM}C, \figref{MAM}B).

\begin{figure}[h!]
\begin{center}
\includegraphics[width=180mm]{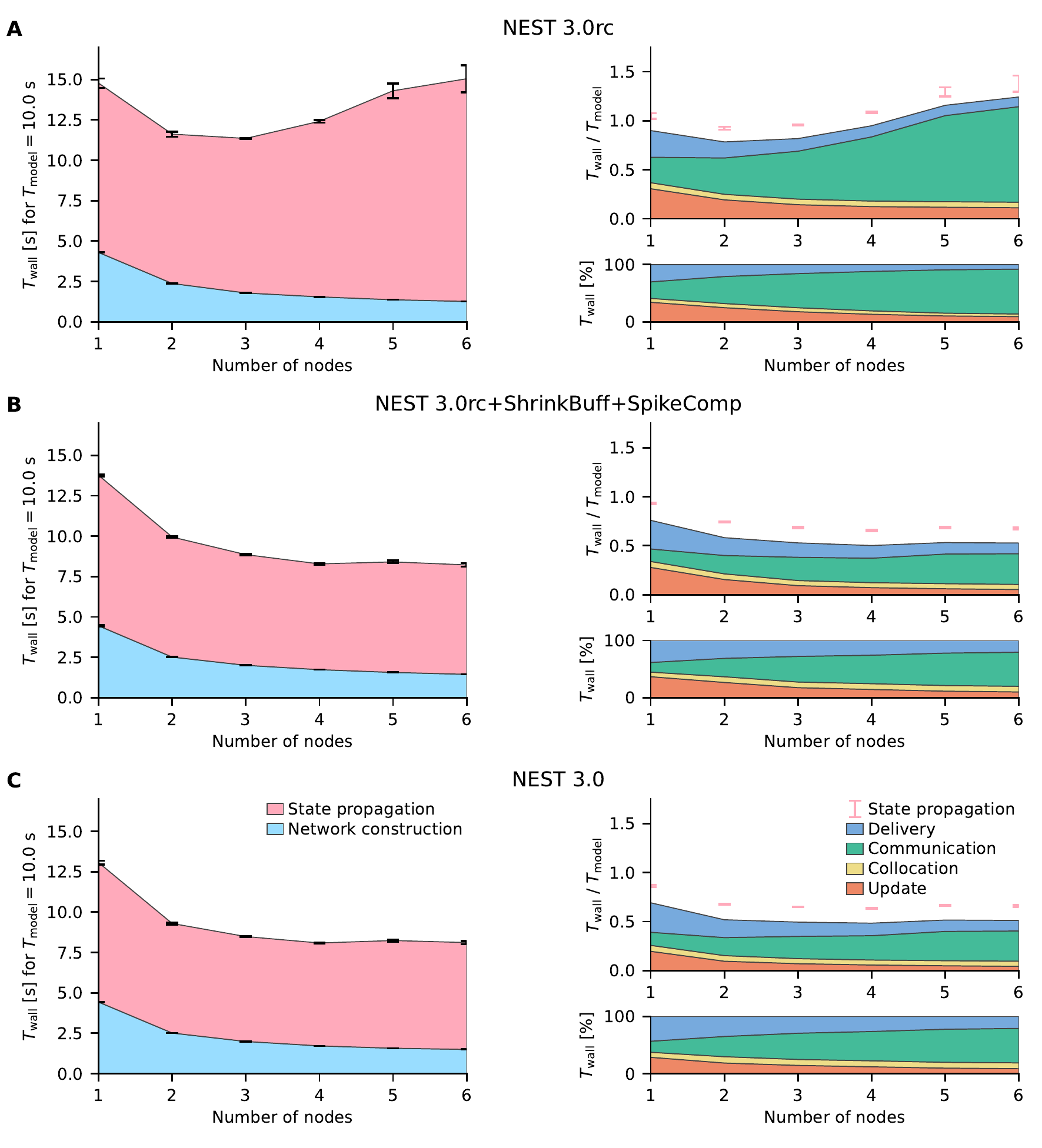}%
\end{center}
\caption{
\textbf{Strong-scaling performance of the microcircuit model on JURECA-DC}.
Same display as in \figref{hpc-benchmark}. The model time is $T_\mathrm{model}=\unit[10]{s}$. Simulations are repeated for ten different random seeds.
\textbf{A}~\texttt{NEST~3.0} release candidate.
\textbf{B}~\texttt{NEST~3.0} release candidate with spike compression and shrinking MPI buffers.
\textbf{C}~\texttt{NEST~3.0}.
\vspace*{1cm}} 
\label{fig:microcircuit}
\end{figure}

Next, we assess the strong-scaling performance of the microcircuit model \citep{Potjans14_785}. The model size is similar to the size of one of the $32$ areas of the multi-area model. The microcircuit therefore requires fewer resources. We show results of the model run on one to six compute nodes; for a detailed analysis of \texttt{NEST}'s thread scaling performance on the example of this model refer to \cite{Kurth21_arxiv}. Using \texttt{NEST~3.0rc}, the microcircuit is simulated faster than the HPC-benchmark and the multi-area models and achieves approximately real time ($T_\text{wall}/T_\text{model}\approx 1$, \figref{microcircuit}A). The finer resolution of the vertical axis of the top-right graph reveals a small gap between the state propagation measured with Python timers and the sum of the phases timed on the C++ level which is not visible for the other models. The state-propagation time of the microcircuit is also dominated by the communication phase similarly to the respective benchmarks with the multi-area model (\figref{MAM}B) and even increases with the number of nodes used. However, shrinking MPI buffers does not reduce communication significantly (data not shown), indicating that we face a different bottleneck with the microcircuit model. With on the order of $10^3$ outgoing connections per neuron, a single neuron of this model has multiple targets on each MPI process and, in particular, on multiple threads of a given process. Since the 5g kernel is designed to send out a separate copy of a neuron's spiking information to each target thread, multiple copies of identical information about the activity of a presynaptic neuron may be sent to the same process, causing unnecessary communication load. To tackle this, we devise a spike compression algorithm which only requires transmitting the spiking information once to each MPI process where it is locally routed to the receiving threads. For details see \secref{methods_spike-compression} ``\nameref{sec:methods_spike-compression}".
This algorithm leads to a significant reduction in communication time for the microcircuit model (compare \figref{microcircuit}A, \figref{microcircuit}B).

The microcircuit model easily fits within the main memory of one compute node of JURECA-DC. Due to the simplicity of the employed model neurons and the absence of synaptic plasticity mechanisms, the network model causes little workload during update and delivery in a strong-scaling experiment\textemdash real-time simulation is already possible with a single compute node. Consequently, communication naturally starts to dominate the state-propagation time at a few compute nodes even with the spike-compression optimization described above. While increasing the number of compute nodes from one to two still results in a fair reduction of state-propagation time, scaling is already sublinear, and increasing the number of compute nodes further hardly results in further improvement. Therefore, simulation phases other than the so far discussed communication become important if the objective of the optimizations is, for example, achieving real-time performance with even fewer resources. In the following we highlight an algorithm adaptation that concentrates on the update phase. A redesign of the neuronal input buffers prevents neurons from retrieving the input values for different channels, for example, excitatory and inhibitory, from separate locations in memory. Thereby, the cache can be better utilized during neuronal updates. Instead of maintaining separate buffers for the input channels as in the original 5g kernel, neurons maintain a single buffer with all inputs for a particular simulation time step stored contiguously in memory. For details see \secref{multi-channel-input-buffer} ``\nameref{sec:multi-channel-input-buffer}".
This adaptation is most effective for network models with short minimum synaptic delays; both the microcircuit and the multi-area model use $\unit[0.1]{ms}$.
\figref{microcircuit}C shows the resulting decrease in update time for few compute nodes.

In summary, the analysis with \texttt{beNNch} exposes the communication phase as a major performance bottleneck in microcircuit and multi-area model simulations with the release candidate \texttt{NEST~3.0rc}. The underlying problem is, however, a different one for each of the two models, and they are rectified with different adaptations to the code: the shrinking MPI buffers (\secref{methods_adaptive-buffers}) improve the performance of the multi-area model while spike compression (\secref{methods_spike-compression}) increases simulation speed of the microcircuit model. Notably, none of the adaptations introduce performance regressions for the respective other model (data not shown). In addition, the update phase is improved by introducing neuronal input buffers with multiple channels (\secref{multi-channel-input-buffer}). Returning to the HPC-benchmark model, \figref{hpc-benchmark}B shows that the kernel adaptations are not detrimental to the originally tested model; the overall state-propagation time is preserved with the final \texttt{NEST~3.0} release. However, the reduced communication and update times here come at the cost of increased delivery times due to an additional indirection introduced with spike compression. Ongoing work targets the delivery phase \citep{pronold2021routing_efficient_cache} and gives a perspective for performance improvements in future \texttt{NEST} releases.
\section{Discussion}
\label{sec:discussion}
Benchmarking studies in the field of neuronal network simulations are often hard to reproduce and compare. To overcome this problem, we propose a unified and modular workflow for defining, running, and analyzing benchmark simulations. We identify five dimensions spanning the space of the benchmarking endeavor, and work out their specific challenges: hardware configuration, software configuration, simulators, models and parameters, and researcher communication. The  benchmarking concept developed in this study encompasses all five dimensions and proposes solutions for the posed challenges in the form of self-contained and interacting modules. Each module contributes to one of the main workflow segments: configuration and preparation, actual benchmarking, data- and metadata handling, and data presentation. As a proof of concept, we provide a reference implementation of the framework (\texttt{beNNch}), describe the concrete underlying technologies, and apply it to a specific use case: assessing and comparing the performance of different versions of the neuronal network simulator \texttt{NEST} for three different network models. The reference implementation goes beyond existing benchmarking environment software such as \texttt{JUBE}: it adds an interface to models, installs and deploys simulation software, automates data and metadata annotation, and implements storage and presentation of results. The use case illustrates how the framework helps to focus simulator development by detecting performance bottlenecks, and demonstrates the relevance of an accessible and comprehensive benchmarking setup. The software is ready to use, not only for developers of simulation technology, but also for researchers seeking to find optimal performance configurations for their models.

The proposed workflow is generic and not restricted to benchmarking neuronal network simulations with \texttt{NEST}. The reference implementation, however, still faces limitations and open problems.
First, it is \textit{a~priori} unclear what parameters, configurations or external influences may possibly contribute to differences in the performance of complex software systems such as simulation engines. \texttt{beNNch} seeks to address this problem by employing a metadata archive which\textemdash in addition to the selection of metadata directly attached to the performance results\textemdash tracks further metadata that are seemingly insignificant at the time of simulation but may become relevant in future investigations. Exhaustiveness, however, can not be claimed. For the exploration and presentation of benchmarking data, the reference implementation uses metadata to filter benchmark results and to highlight differences in a flip-book format. However, even if all relevant metadata were tracked, selecting sensible metadata keys for filtering and highlighting is a hard problem. In the current implementation, this requires knowledge about existing results and, therefore, human input. Future solutions could, for example, categorize and hierarchically structure metadata keys to facilitate and semi-automatize these steps.
Second, the network model specifications, expressed in the \texttt{PyNEST} set of commands for the Python language, require adaptations to interface with the benchmarking framework. These include accepting parameters passed by \texttt{JUBE} benchmarking files, adjusting the model specification to work with different versions of the simulation engine, and storing recorded metadata and performance measures such as the duration of simulation phases. At the moment, it is a manual task to keep the benchmarking model version up to date with the original model version, which is error prone. We use rigorous version control of the code, automatic checking for errors (via exceptions), and continuous testing for correct simulation outcome to reduce the risk of errors. This strategy could be automatized further in the future by finding methods to automatically inject respective instrumentation into the executable model descriptions. To mitigate the additional overhead, we keep the necessary changes as minimal as possible, thereby lowering the entry barrier for new models.
Third, the reference implementation makes concrete choices on the employed tools. Alternatives, however, may be viable. 
For example, the required software for the simulations is installed with \texttt{Builder} which can be integrated with other package management systems or replaced by a different solution. Our strategy exploits the native software environment available on a compute cluster which is typically specifically configured for the underlying hardware. An alternative is to use containerized systems such as \texttt{Docker}\footnote{\url{https://www.docker.com}} or \texttt{Singularity}\footnote{\url{https://sylabs.io}}.
Replacing \texttt{NEST} by a different simulator requires adapting the model implementations. Expressing the models in the simulator-independent language \texttt{PyNN} \citep{Davison09_10} instead of \texttt{PyNEST} would avoid this. However, additional layers of complexity such as \texttt{PyNN} may have an impact on performance, making it more challenging to pinpoint bottlenecks in the simulator backend. \texttt{JUBE} as an environment to manage jobs on compute clusters could be substituted by tools such as  \texttt{ecFlow}\footnote{\url{https://confluence.ecmwf.int/display/ECFLOW}} or \texttt{AiiDA}\footnote{\url{https://www.aiida.net}}.
Further, one could replace \texttt{git-annex} with, e.g., \texttt{DataLad}\footnote{\url{https://www.datalad.org}} which is based on the same technology but extends its functionality and provides slightly different metadata handling. The flip-book-style presentation of results could also be replaced or supplemented with other approaches, for example an automatically generated overview figure showing results from multiple benchmarking runs together, similar to Figures \ref{fig:hpc-benchmark}--\ref{fig:microcircuit} in this paper.
Fourth, \texttt{beNNch} presently focuses on a single performance measure: the time-to-solution. However, different performance aspects, such as energy-to-solution and memory consumption, may also be of interest. Energy-to-solution, for example, combines power consumption and time-to-solution. Monitoring both power consumption and time-to-solution enables researchers to determine an optimal number of compute nodes balancing speed and energy consumption \citep{VanAlbada18_291}.
The memory consumption of the simulation dictates, for instance, the smallest number of nodes required to simulate a network of a given size, or the largest network size possible to simulate on a given machine. Reducing memory requirements was a major driving force behind the improvements to the \texttt{NEST} kernel \citep{Helias12_26,Kunkel2012_5_35,Kunkel14_78,Jordan18_2} in the past decade. The spike compression introduced here reduces the time-to-solution (communication phase, Figures \ref{fig:hpc-benchmark} and \ref{fig:microcircuit}). However, this code change directly affects the memory consumption. Assuming that the number of postsynaptic targets per neuron is fixed, the memory overhead is negligible if the number of MPI processes is small. But in the limit of a large number of MPI processes, i.e., when each neuron has at most one target on each process, the effective size of each synapse is increased by $8$ byte. In this limit, users thus are encouraged to actively deactivate the ``spike compression" feature. This example illustrates that performance optimizations often have to find a balance between acceptable solutions for different measures. Due to its modular structure, \texttt{beNNch} is ready to include further performance measures.

To achieve long term sustainability, organized and openly available communication on development is essential. Adhering to this guideline, we have developed \texttt{beNNch} as an open source software project from the start, making use of a public issue tracker, suggestions via pull requests, public code reviews, and detailed documentation. This approach facilitates constructive communication between users and developers which enables a targeted progression of the framework by demand. While the concrete application of \texttt{NEST} benchmarks of neuronal network models shaped our specific implementations, the modular structure allows for adaptation to other use cases.
In certain domains of software development, it is already common practice to verify each code change on the basis of syntax, results, and other unit tests. The proposed automated approach to execute performance benchmarks creates the opportunity to integrate an aspect of validation  directly into the development cycle. This way, performance regressions of algorithm adaptations are immediately exposed, while positive effects can readily be demonstrated. For high-performance software, however, comprehensive checks for scaling performance are particularly costly because they require compute time on state-of-the-art clusters and supercomputers. Therefore, it is important to conduct the performance benchmarks purposefully and with care. By organizing benchmarking results and keeping track of metadata, \texttt{beNNch} helps to avoid redundant benchmark repeats and instead encourages a direct comparison with previous results.

It has long been recognized that software development in science underlies different constraints and needs to fulfill different requirements as compared to industry \citep{Diesmann01}. The software crisis in neuroscience at the beginning of the century led to the foundation of the International Neuroinformatics Coordinating Facility (INCF) in 2005. A first INCF report in 2006 addresses the software challenges of large-scale modeling in neuroscience \citep{INCFsecretariat18} and recommends establishing a common set of benchmark models and a corresponding framework for assessing accuracy and efficiency. Furthermore, the report advocates benchmarking neuroscientifically relevant published models rather than network models constructed specifically for the purpose of benchmarking only. In 2007, the community made a first effort in verifying simulation codes by using a number of simple network models \citep{Brette07_349}. Executable model descriptions are, in part, already expressed in the simulator independent language PyNN \citep{Davison09_10}, but there is no support by a common benchmarking framework, and a focus is set on correctness rather than computational efficiency. The emerging field of Research Software Engineering (RSE) is studying how, in the scientific setting, reliable and sustainable software can be developed, developers can be educated for this purpose, and science organizations and politics can be made aware of its strategic relevance \citep[Manifesto\footnote{\url{https://www.software.ac.uk/about/manifesto}} and][]{Akhmerov19_572}. Obvious differences to software engineering in the industrial setting include research code being developed by scientists rather than experienced software developers, the time-constrained and thesis-bound nature of scientific projects, and the continuous integration of new contributors into the development process.
Our study contributes to RSE conceptually by identifying the dimensions of benchmarking simulation technology and proposing a general workflow capable of coping with the complexity, and practically by developing a reference implementation of a benchmarking framework which can be used to test and refine the concepts. It is too early to tell quantitatively whether the benchmarking framework improves the collaboration in a joint project and the communication between researchers in the community.

The proposed framework enables benchmarking of research software to evolve from one-off tasks of individual researchers to a collaborative routine effort, thereby increasing the benchmarking capacity and reducing its susceptibility to errors. Making \texttt{beNNch} accessible to the community as an open-source software puts the concept to the test. We are looking forward to learn how the current implementation of the framework's components are received and adapted to other applications. Due to the conceptual foundation and modular structure, we hope that \texttt{beNNch} can adjust to future requirements and ultimately help increase the complexity and explanatory scope of brain models. The benchmarking concepts developed in this work are not limited to neuroscience and can be transferred to other types of simulation research.

\section{Materials and Methods}
\label{sec:materials_and_methods}
\subsection{NEST developments}
\label{sec:methods_nest-developments}

\subsubsection{Shrinking MPI buffers}
\label{sec:methods_adaptive-buffers}

Motivated by reducing the memory footprint of the postsynaptic infrastructure\textemdash necessary to deliver spikes to their process-local targets\textemdash the 5g kernel of \texttt{NEST 3.0rc} prepares a separate part of the MPI send buffer for each target process and only includes the relevant spikes.
Thus, each process is responsible for sending the spikes of its neurons to all target processes for each communication time step.
\texttt{NEST 3.0rc} implements a homogeneous buffer size across processes to avoid overhead introduced by variable buffer sizes; in the latter case, each process would need to complete two rounds of communication, one for transmitting the size, and one for the actual spiking information.
Similarly, transmitting a certain amount of information via sending MPI buffers is more efficient when fewer buffers\textemdash each carrying more information\textemdash are sent.
\texttt{NEST 3.0rc} consequently aims to reduce the number of needed MPI buffers to only $1$ by dynamically increasing the global buffer size whenever a process cannot fit all spikes into the buffer.
Specifically, every time more than a single buffer needs to be sent by a process, \texttt{NEST} increases the buffer size of the following communication step by a factor of $1.5$.
In this scheme, a reduction of buffer sizes is not implemented, meaning that buffer sizes can only increase or stay constant.
The kernel of \texttt{NEST 3.0rc+ShrinkBuff} addresses this by introducing the following algorithm for shrinking the global buffer size.
In each communication round in which only a single send buffer is required, the buffer for the following round decreases by a factor of $1.1$.
Even though this implementation leads to an oscillation of buffer size for constant spiking activity, tests show that this simple mechanism only introduces negligible cost while being robust.

\subsubsection{Spike compression}
\label{sec:methods_spike-compression}

\begin{figure}[t]
\begin{center}
\includegraphics[width=85mm]{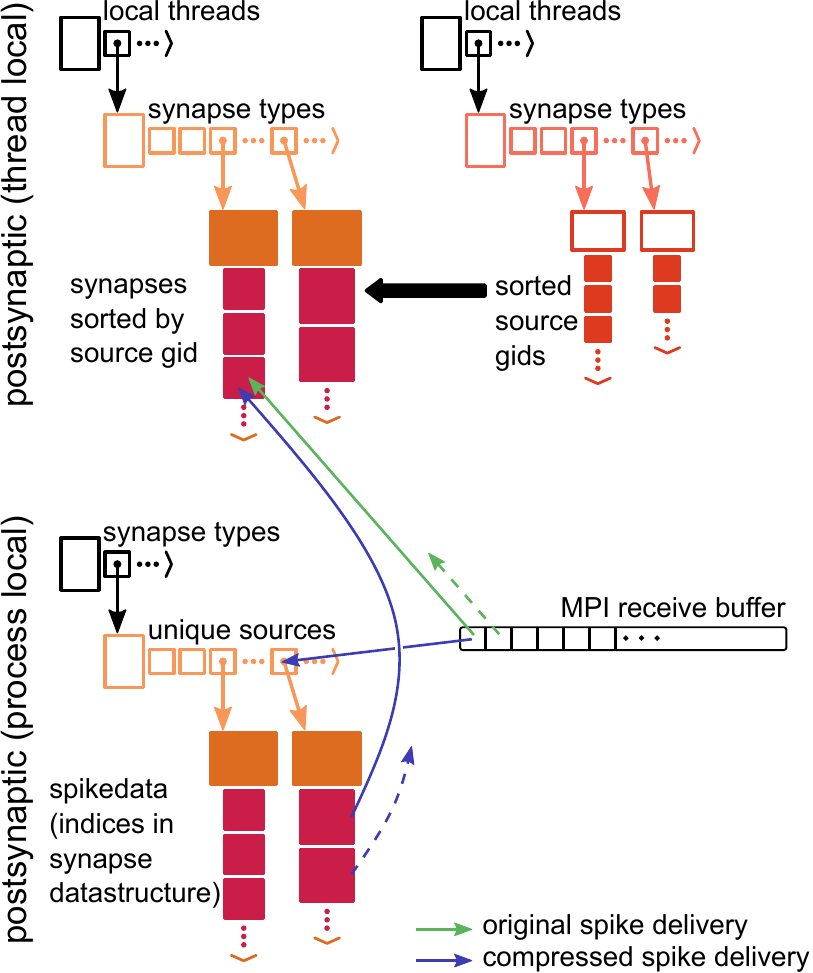}%
\end{center}
\caption{
\textbf{Spike compression adds an additional indirection to postsynaptic spike routing.}
Green arrow denotes original spike delivery introduced with the 5g kernel \citep[][same display as their Figure 4A]{Jordan18_2}.
Blue arrow illustrates additional indirection with compressed spike delivery.
Dashed arrows indicate spikes from the same source neuron with target on a different thread.
}
\label{fig:infrastructure-5g-compressed-spikes}
\end{figure}

\texttt{NEST}'s 5g kernel \citep{Jordan18_2} introduces a two-tier connection infrastructure for routing spikes.
The connection infrastructure consists of data structures on the presynaptic side (the MPI process of the sending neuron) and the postsynaptic side (the MPI process of the receiving neuron), cf. \secref{instrumentation}.
Communication of spikes is organized as follows: when a neuron becomes active, its targets are retrieved from the local presynaptic data structure.
These targets represent indices of synapses in the ``thread-local" postsynaptic data structure through which spikes are routed to the target neurons.
The presynaptic side then creates MPI buffers containing collections of such indices which are subsequently communicated to the postsynaptic side via the MPI \texttt{Alltoall} function.
To deliver spikes on the postsynaptic side, each thread uses the received spikes to index its local postsynaptic data structure and register a spike in the corresponding synapse (\figref{infrastructure-5g-compressed-spikes}, ``original spike delivery").
If a presynaptic neuron has targets on multiple threads of a process, it hence has to send multiple spikes, i.e., indices in different thread-local data structures, to the target process.

Here, we adapt this infrastructure as follows.
We introduce an additional data structure on the postsynaptic side which is shared across threads (``process local").
This data structure contains, arranged by source neuron, the indices of all process-local synapses.
While the presynaptic part of communicating spikes remains essentially identical, the postsynaptic part incurs an additional indirection: Each entry in the MPI receive buffer now represents an index in the new process-local postsynaptic data structure.
Using this index, each thread can retrieve the indices of thread-local targets, to which it can then deliver spikes as previously (\figref{infrastructure-5g-compressed-spikes}, ``compressed spike delivery"; note that the origin of the dashed arrow changes).
In contrast to the previous implementation, each presynaptic neuron thus sends at most one spike to each process.

In \texttt{NEST 3.0}, spike compression is turned on by default, but the previous 5g behavior can be recovered by setting:
\lstset{language=myPython}
\begin{nolinenumbers}
\begin{lstlisting}
nest.SetKernelStatus({"use_compressed_spikes": False})
\end{lstlisting}
\end{nolinenumbers}

\subsubsection{Neuronal input buffers with multiple channels}
\label{sec:multi-channel-input-buffer}
\begin{figure}[h!]
\begin{center}
\includegraphics[width=85mm]{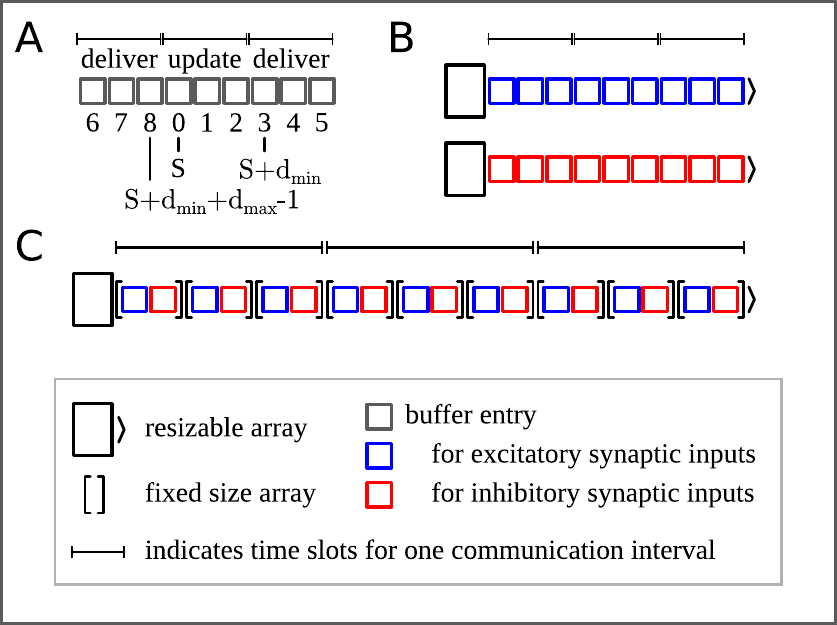}%
\end{center}
\caption{
\textbf{Neuronal input buffers accounting for synaptic delays in simulations of spiking neuronal networks.} \textbf{A}~Structure of neuronal input buffers assuming a minimum synaptic delay $d_\mathrm{min}$ of three simulation time steps and a maximum delay $d_\mathrm{max}=2d_\mathrm{min}$. To buffer upcoming inputs during simulation a total buffer size of $d_\mathrm{min}+d_\mathrm{max}$ time slots is required, which corresponds to three communication intervals of three simulation time steps each. After every spike communication and subsequent spike delivery to local targets, simulation time is advanced, meaning that the relative time origin $S$ of the neuronal input buffers advances by $d_\mathrm{min}$ time slots with a wrap-around at the buffer end. A pre-calculated and continuously updated look-up table maps the index relative to $S$ to the actual buffer index. Example: The relative time origin $S$ is located at the fourth time slot. Synaptic delays of the inputs of the middle buffer segment elapse with the upcoming three simulation time steps; the neuron integrates these inputs updating its state. Spikes are then communicated and new inputs delivered to the neuron are added to the time slots in the last or first buffer segment depending on the delay, which is at least $d_\mathrm{min}$ and at most $d_\mathrm{max}$. Relative time origin $S$ then advances to the seventh buffer slot (not shown).
\textbf{B}~Original neuronal spike buffers for two input channels (e.g., excitatory and inhibitory synaptic inputs). For each channel a separate resizable array buffers the inputs for the upcoming time slots.
\textbf{C}~Multi-channel input buffer for two input channels. A single resizable array stores the inputs for the upcoming time slots, where for each time slot a fixed size array holds the inputs sorted by channel.}
\label{fig:multi-channel-input-buffer}
\end{figure}

Simulation technology for spiking neuronal networks requires techniques to handle synaptic transmission delays. The reference simulation code (\secref{NEST}) follows a globally time-driven approach: spikes are constrained to a time grid and regularly exchanged between MPI processes using collective communication. The time grid defines the simulation time step for neuronal updates, whereas the minimum synaptic delay $d_\mathrm{min}$ in the network model defines the communication interval \citep{Morrison05}, which comprises at least one simulation time step. In the microcircuit model and the multi-area model used in this study the minimum delay is $\unit[0.1]{ms}$ (i.e., $d_\mathrm{min}=1$ simulation time step) and in the HPC-benchmark model it is $\unit[1.5]{ms}$ (i.e., $d_\mathrm{min}=15$ simulation time steps). While communication and subsequent process-local delivery of spikes define interaction points between neurons, within a communication interval each neuron independently updates its state for all time steps without interruption. Hence, a simulation cycle of neuronal update, spike-communication, and spike-delivery phase propagates the network state by one communication interval, but within each update phase neurons propagate their state in potentially shorter simulation time steps.
All spikes emitted by the process-local neurons during such an update are immediately transmitted during the subsequent communication and on the receiver side delivered to their target neurons. Hence, to account for synaptic delays, neurons cannot immediately integrate the incoming spikes into their dynamics, but they need to buffer the inputs until the corresponding delays elapse. To this end, neurons maintain input buffers of $d_\mathrm{min}+d_\mathrm{max}$ time slots, where $d_\mathrm{max}$ denotes the maximum synaptic delay in the network (\figref{multi-channel-input-buffer}A). The relative time origin $S$ defining the time slots from which to retrieve inputs during update and the time slots for adding inputs during spike delivery advances by $d_\mathrm{min}$ time slots at the end of every simulation cycle. In this way, the time slots that were read and reset during the update of the current cycle become available for adding new inputs during the spike delivery in the next cycle.
For cases where the communication interval comprises multiple simulation time steps (e.g., HPC-benchmark model), input retrieval is most costly for the first step as the corresponding buffer entry needs to be loaded into cache, but then benefits from the already cached subsequent buffer entries in the subsequent steps of the communication interval. If, however, the communication interval consists of only one simulation step due to a very short minimal synaptic delay (e.g., microcircuit and multi-area model), input retrieval is costly for every simulation step as each step is handled in a separate simulation cycle, and hence caching of relevant input buffer entries is rendered ineffective during the spike communication and delivery that follows each neuronal update phase.

Most neuron models need to distinguish between input channels to treat the corresponding inputs dynamically differently, as for example, excitatory and inhibitory synaptic inputs causing different postsynaptic responses. The original input-buffer design required a separate resizable array per channel storing the channel's input values per time slot (\figref{multi-channel-input-buffer}B). This entailed retrieval of the input values for a particular time step from separate locations in memory, which amplifies the cache inefficiency during update for network models with short minimum delays described above.
To alleviate this issue, the newly introduced input buffer allows storing the input values for multiple channels per time slot contiguously in fixed size arrays in a single resizable array (\figref{multi-channel-input-buffer}C). Thus, neurons now retrieve all input values for a particular time step by accessing subsequent locations in memory in one pass.
\section*{Conflict of Interest Statement}
The authors declare that the research was conducted in the absence of any commercial or financial relationships that could be construed as a potential conflict of interest.

\section*{Author Contributions}
Study design: JA, JP, ACK, SBV, KHM, AP, DT, TT, MD, JS\\
Implementation of \texttt{beNNch}: JA, JP, ACK, SBV, KHM, DT, JS\\
Execution and analysis of benchmarks: JA\\
Figures: JA, JP, SK, JJ\\
Implementation of shrinking MPI buffers: AP, JA\\
Implementation of spike compression: JJ, JS\\
Implementation of neuronal input buffers with multiple channels: SK\\
All authors contributed to the writing of the manuscript and approved it for publication.

\section*{Funding}
This project has received funding from
the European Union’s Horizon 2020 Framework Programme for Research and Innovation under Specific Grant Agreement No. 945539 (Human Brain Project SGA3) and no. 754304 (DEEP-EST);
the Helmholtz Association Initiative and Networking Fund under project number SO-092 (Advanced Computing Architectures, ACA);
the Joint Lab ``Supercomputing and Modeling for the Human Brain";
the Deutsche Forschungsgemeinschaft (DFG, German Research Foundation) - 368482240/GRK2416;
and
the Helmholtz Metadata Collaboration (HMC), an incubator- platform of the Helmholtz Association within the framework of the Information and Data Science strategic initiative, under the funding ZT-I-PF-3-026.

\section*{Acknowledgments}
We thank the members of the NEST development community for their contributions to the concepts and implementation of the NEST simulator,
and our colleagues in the Simulation and Data Laboratory Neuroscience of the Jülich Supercomputing Centre for continuous collaboration.
We gratefully acknowledge the computing time granted by the JARA Vergabegremium and provided on the JARA Partition part of the supercomputer JURECA at Forschungszentrum Jülich (computation grant JINB33).
We acknowledge the use of Fenix Infrastructure resources, which are partially funded from the European Union's Horizon 2020 research and innovation programme through the ICEI project under the grant agreement No. 800858.

\section*{Data Availability Statement}
The benchmarking framework is publicly available under \url{https://github.com/INM-6/beNNch}.
The data sets generated and analyzed for this study as well as the code to reproduce all figures of this paper is available at Zenodo (\url{https://doi.org/10.5281/zenodo.5784634}).
An exemplary flip-book containing the results shown in this work can be accessed under \url{https://inm-6.github.io/beNNch}.


\end{document}